\pdfoutput=1

\documentclass[11pt]{article}

\usepackage[]{emnlp2021}

\usepackage{times}
\usepackage{latexsym}
\usepackage{microtype}
\usepackage{graphicx}
\usepackage{booktabs}  
\usepackage{multirow}
\usepackage{subfigure}
\usepackage{verbatim}
\usepackage{CJKutf8} 
\usepackage{amssymb}
\usepackage{amsmath}

\usepackage{algorithm}
\usepackage{algorithmic}
\usepackage{makecell}
\usepackage{flushend}
\usepackage[T1]{fontenc}

\usepackage[utf8]{inputenc}

\usepackage{microtype}

\title{Are Big Recommendation Models Fair to Cold Users?}

\author{Chuhan Wu$^\dagger$~~~~Fangzhao Wu$^\ddagger$~~~~Tao Qi$^\dagger$~~~~\textbf{Yongfeng Huang}$^\dagger$\\
    $^\dagger$Department of Electronic Engineering, Tsinghua University, Beijing 100084, China  \\
     $^\ddagger$Microsoft Research Asia, Beijing 100080, China\\
  \tt{\{wuchuhan15,wufangzhao,taoqi.qt\}@gmail.com,}\\
  \tt{yfhuang@tsinghua.edu.cn}
  }

\begin{document}
\maketitle
\begin{abstract}

Big models are widely used by online recommender systems to boost recommendation performance.
They are usually learned on historical user behavior data to infer user interest and predict future user behaviors (e.g., clicks).
In fact, the behaviors of heavy users with more historical behaviors can usually provide richer clues than cold users in interest modeling and future behavior prediction.
Big models may favor heavy users by learning more from their behavior patterns and bring unfairness to cold users.
In this paper, we study whether big recommendation models are fair to cold users.
We empirically demonstrate that optimizing the overall performance of big recommendation models may lead to unfairness to cold users in terms of performance degradation.
To solve this problem, we propose a \textit{BigFair} method based on self-distillation, which uses the model predictions on original user data as a teacher to regularize predictions on augmented data with randomly dropped user behaviors, which can encourage the model to fairly capture interest distributions of heavy and cold users. 
Experiments on two datasets show that \textit{BigFair} can effectively improve the performance fairness of big recommendation models on cold users without harming the performance on heavy users.
\end{abstract}

\section{Introduction}

With the explosion of online user data and the increase of computing resources, big models are widely used in many personalized recommendation scenarios such as news recommendation~\cite{wu2021empowering} and video recommendation~\cite{covington2016deep}.
For example, \citet{ma2018modeling} proposed a multi-task learning method with multi-gate mixture-of-experts to train large-scale recommendation models in multiple tasks.
\citet{wu2021empowering} proposed a pretrained language model (PLM) empowered news recommendation method that uses PLMs to model deep semantic information of news.
These powerful big recommendation models have achieved notable performance gains in various scenarios by mining rich information from a huge volume of training data~\cite{naumov2019deep,sun2019bert4rec,zhang2021unbert,jia2021rmbert}.

\begin{figure}[!t]
  \centering 
      \includegraphics[width=0.99\linewidth]{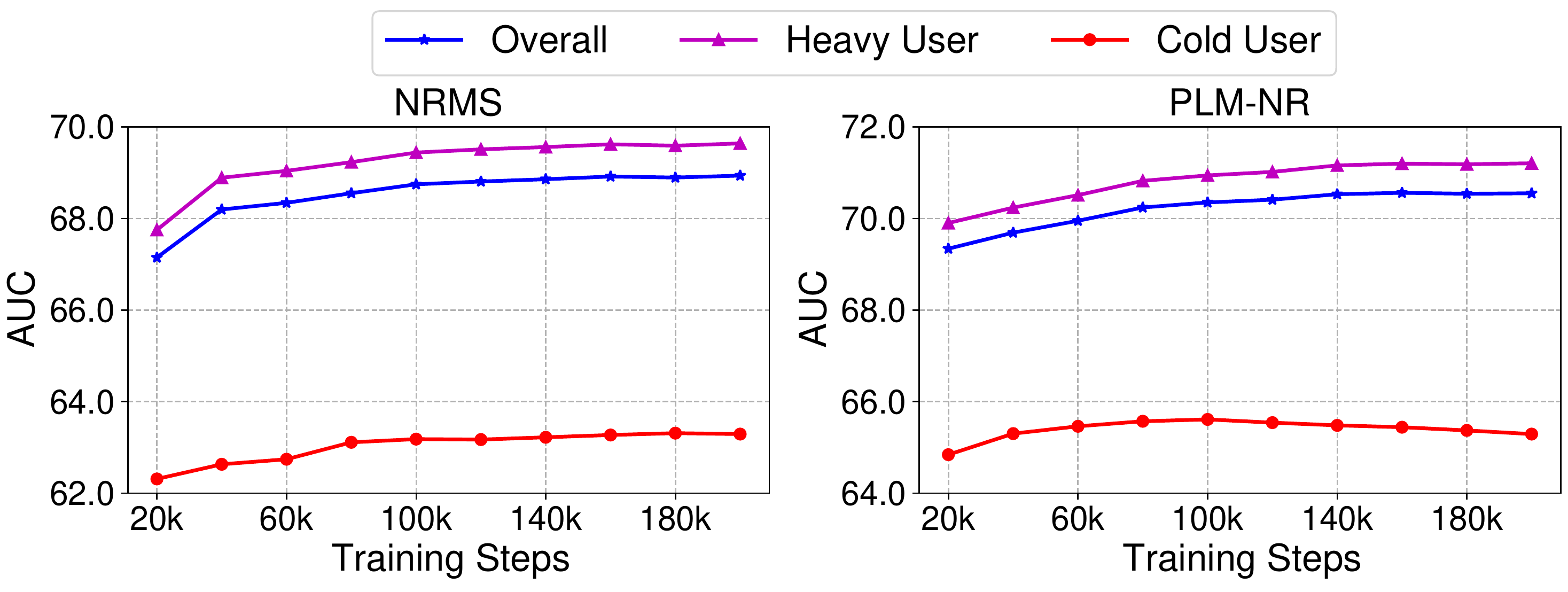} 
  \caption{Performance of small and big recommendation models on cold and heavy users after different training steps on the MIND dataset.}\label{fig.exp}  
\end{figure}

In the recommendation scenario, heavy users (aka active users) usually have much more behaviors than cold users, and thereby can provide richer clues on inferring user interest  and predicting future behaviors~\cite{li2021user}.
Due to the high capacity of big recommendation models, they may have higher risks than small models in tendentiously fitting the rich patterns encoded in heavy users' behaviors rather than cold users. 
Fig.~\ref{fig.exp} shows an example of this phenomenon.
We compare the news recommendation performance of a small model NRMS~\cite{wu2019nrms} based on shallow self-attention networks and a big model PLM-NR~\cite{wu2021empowering} based on PLMs after different training steps on the MIND~\cite{wu2020mind} dataset.\footnote{Following~\cite{he2016ups}, users with no more than 5 historical behaviors are regarded as cold users.}
We find that the small model's performance on cold and heavy users consistently improves during model training.
On the contrary, the performance of big model  only continuously improves on heavy users, while the performance on cold users starts to decline after around 100k steps.
This shows that simply optimizing the overall performance of big recommendation models may lead to a sacrifice of cold users' experience, which is unfair to them.

In this paper, we investigate whether big recommendation models bring unfairness to cold users.
Through extensive experiments on two real-world datasets, we empirically demonstrate that optimizing the overall accuracy of big recommendation models may lead to unfairness to cold users in terms of recommendation performance degradation.
In addition, we propose a \textit{BigFair} method with a self-distillation framework to solve this problem.
It uses model predictions on original user data as a virtual teacher to regularize the model prediction on augmented user data with randomly dropped historical behaviors, which can encourage the model to fairly capture interest distributions of heavy and cold users.
Experimental results on both datasets show that \textit{BigFair} can improve the fairness of big recommendation models to cold users without hurting the performance on heavy users.

\section{Performance Unfairness of Big Recommendation Models}\label{sec:Model}

In this paper, we take personalized news recommendation as an example scenario since it can naturally incorporate big pre-trained language models (e.g., BERT) for news text modeling to form a big recommendation model.
The task of news recommendation can be formulated as predicting whether a target user $u$ with a list of historical clicked news $[D_1, D_2, ..., D_N]$ ($N$ is the history length) will click on a candidate news $D_c$.
We take the PLM-based news recommendation framework proposed in~\cite{wu2021empowering} for our experiments.
The datasets, experimental settings and results are introduced in the following sections.

\subsection{Datasets and Experimental Settings}

We conduct experiments on two real-world news recommendation datasets.
The first one is the \textit{MIND} dataset~\cite{wu2020mind}, which contains news click data of 1 million users during six weeks.
The second one is collected by ourselves from a commercial online news platform (denoted as \textit{Industry}) from 06/24/2021 to 08/10/2021.
On both datasets, click logs in the last week are used for test, while the rest are for training and validation.
The statistics of them are shown in Table~\ref{dataset}.

\begin{table}[!t]
\centering
\resizebox{1\linewidth}{!}{ 
\begin{tabular}{lcc}
\Xhline{1.5pt}
                     & \textit{MIND}       & \textit{Industry}       \\ \hline
\#users              & 1,000,000  & 18,343,725   \\
\#news               & 161,013    & 1,798,942  \\
\#impressions        & 2,232,748  & 46,781,657  \\
\#clicks             & 3,383,656  & 80,796,599 \\
avg. \#tokens per title & 13.64      & 17.94      \\
avg. history len.    & 36.72      & 226.49      \\
ratio of cold users  & 0.1108      & 0.0869      \\
\Xhline{1.5pt}
\end{tabular}
}
\caption{Statistics of \textit{MIND} and \textit{Industry} datasets.}\label{dataset}
\end{table}

In our experiments, we use Adam~\cite{kingma2014adam} as the optimizer.
The batch size is 32 on \textit{MIND} and 128 on \textit{Industry}.
We regard users with $\leq5$ historical news clicks as cold users.
For all methods, we use the InfoNCE~\cite{oord2018representation} loss for model training.
Following~\cite{wu2019npa}, the number of negative samples associated with each positive sample is 4.
We use AUC as the metric, and we use the performance differences on cold users between the model with optimal overall results and the optimal model on cold users to measure unfairness.
We repeat each experiment 5 times and report the average scores.

\begin{table*}[!t]
\centering
\begin{tabular}{lccccccc}
\Xhline{1.0pt}
\multicolumn{1}{c}{\multirow{2}{*}{\textbf{Model}}} & \multicolumn{3}{c}{\textbf{Model Best for All Users}} & \multicolumn{3}{c}{\textbf{Model Best for Cold Users}} & \multirow{2}{*}{\textbf{Unfairness}} \\
\multicolumn{1}{c}{}                       & Overall        & Heavy        & Cold        & Overall         & Heavy        & Cold         &                             \\ \hline
NAML                                       & 67.89          & 68.53         & 62.75       & 67.86           & 68.49         & 62.78        & 0.03                        \\
LSTUR                                      & 68.45          & 69.12         & 63.10       & 68.42           & 69.08         & 63.16        & 0.06                        \\
NRMS                                       & 68.94          & 69.64         & 63.29       & 68.89           & 69.58         & 63.34        & 0.05                        \\ \hline
PLM-NR (BERT)                              & 69.55          & 70.21         & 64.28       & 69.38           & 69.97         & 64.63        & 0.35                        \\
PLM-NR (RoBERTa)                           & 69.58          & 70.23         & 64.37       & 69.42           & 70.00         & 64.74        & 0.37                        \\
PLM-NR (UniLM)                             & 70.57          & 71.23         & 65.32       & 70.37           & 70.96         & 65.64        & 0.32                        \\ 
\Xhline{1.0pt}
\end{tabular}
\caption{Model results on \textit{MIND}. The checkpoints that are optimal on all users or cold users are evaluated.}\label{result1}
\end{table*}

\subsection{Experimental Results}

\begin{figure}[!t]
  \centering 
      \includegraphics[width=0.75\linewidth]{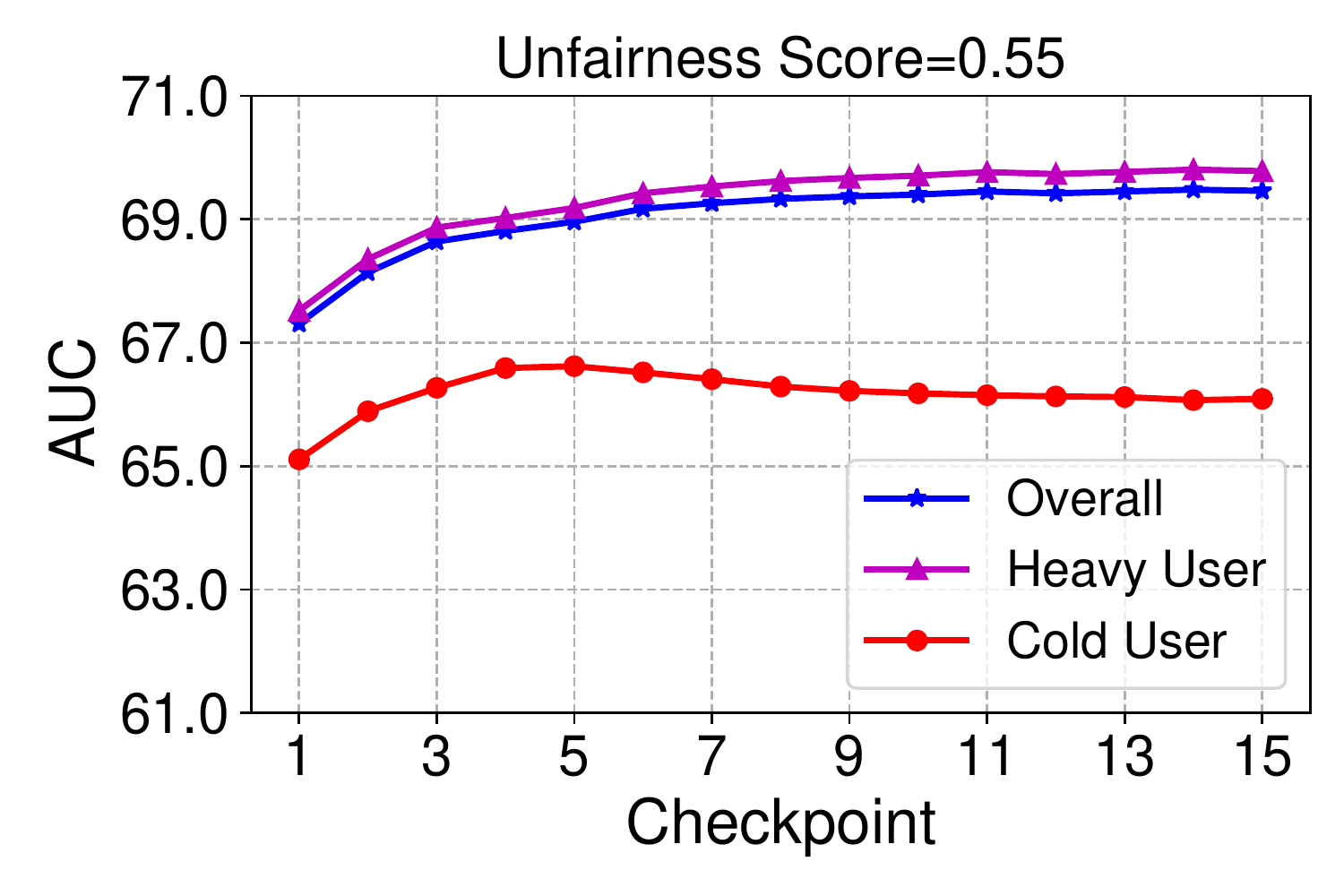} 
  \caption{Results on the \textit{Industry} dataset.}\label{result2}  
\end{figure}

On the \textit{MIND} dataset, we use three widely used small recommendation models for comparison, including: (1) NAML~\cite{wu2019}, using CNN for news modeling and attention network for user modeling; LSTUR~\cite{an2019neural}, a long short-term user modeling method for news recommendation based on GRU network and user ID embedding;
NRMS~\cite{wu2019nrms}, using multi-head self-attention networks for news and user modeling in news recommendation.
The number of parameters in these methods is about 18M\footnote{The top model is rather small because there are more than 17M word embedding parameters}. 
For big recommendation models, following~\cite{wu2021empowering}, we compare the results of using BERT~\cite{devlin2019bert}, RoBERTa~\cite{liu2019roberta} and UniLM~\cite{bao2020unilmv2} models as the news model and an attention network as user model.
These models contain about 111M parameters.
The performance of the optimal checkpoints on all users or cold users as well as their performance differences on cold users (unfairness) are shown in Table~\ref{result1}.
We find the unfairness scores of small models are usually very small, which means that the recommendation accuracy on both heavy and cold users usually improves consistently.\footnote{The unfairness scores of small models can be non-zero because the optimal checkpoints on all users may not always be optimal on cold users.} 
However, for big models there exists a notable cold user performance gap between the checkpoints with the best overall performance and  the cold user performance, which shows that simply optimizing the overall recommendation performance of big recommendation models may bring unfairness to cold users.
It may be because the behavior patterns of heavy users are richer and easier to be captured, and big recommendation models may favor them by optimizing  more on the supervision signals from heavy users.
To further verify the unfairness of big recommendation models, we also conduct experiments on our \textit{Industry} dataset.\footnote{Only the best performed PLM-NR (UniLM) model is evaluated since this dataset is extremely large.}
The results of different checkpoints of the UniLM-based PLM-NR model on \textit{Industry} are shown in Fig.~\ref{result2}.
We can also observe a dramatic performance decline on cold users when optimizing the overall performance,  and the unfairness score reaches 0.55.
It implies severe unfairness issues to cold users.
Thus, it is important to address this defect of big recommendation models to provide better personalized services for cold users.

\section{BigFair}

\begin{figure}[!t]
  \centering 
      \includegraphics[width=0.9\linewidth]{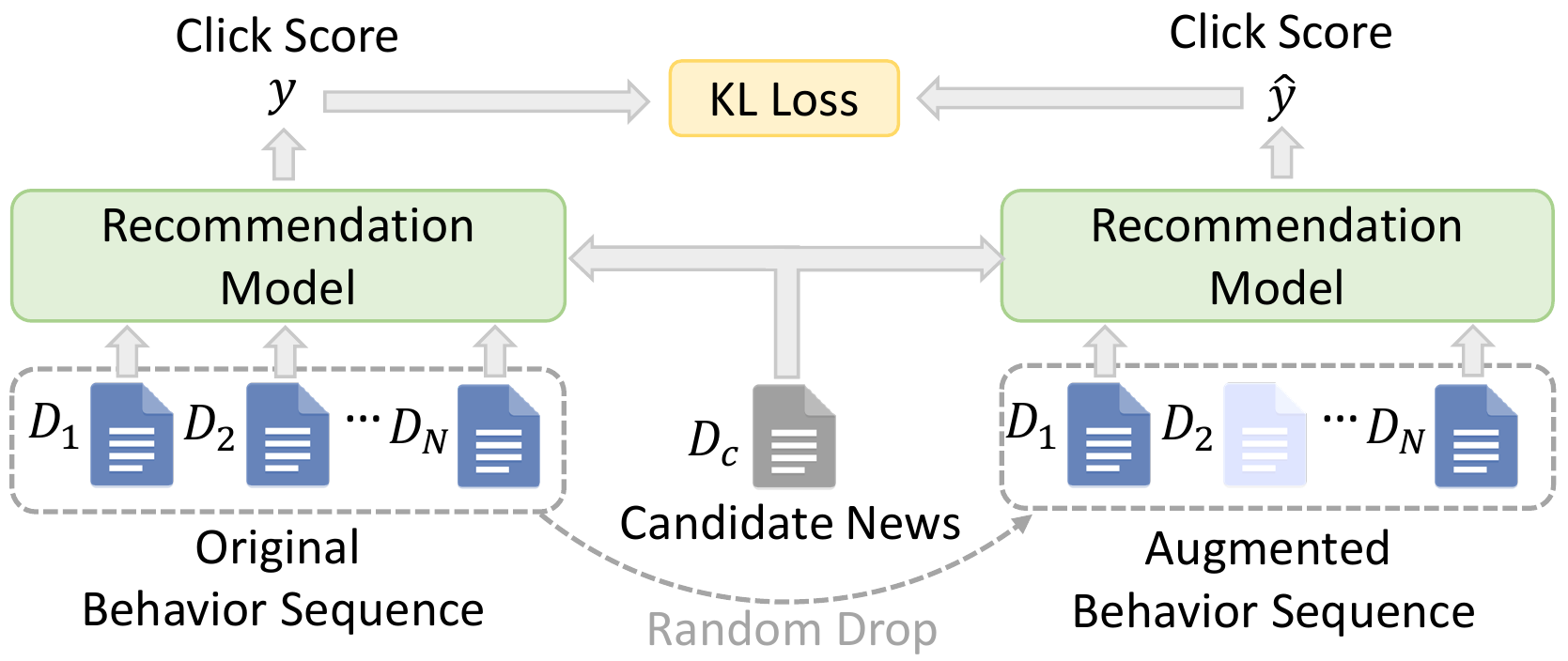} 
  \caption{The self-distillation framework of \textit{BigFair}.}\label{model}  
\end{figure}

To solve the unfairness problem of big recommendation models to cold users, we propose a \textit{BigFair} method based on a self-distillation~\cite{xu2019data} framework, as shown in Fig.~\ref{model}.
Given the behavior sequence of a user, we randomly drop a certain ratio $P$ of its behaviors to form an augmented behavior sequence. 
Both behavior sequences are incorporated by the recommendation model to capture user interest and match candidate news for click score prediction.
Since the behaviors are uniformly dropped, the distribution of user interest encoded in the augmented behavior sequence is expected to be consistent with the original behavior sequence.
Thus, the click scores predicted on augmented user data  are expected to have similar distributions to those predicted on original data.
Motivated by these observations, we propose a self-distillation method by using the click scores predicted on original data as a virtual ``teacher'' to regularize predictions inferred on augmented data.
We apply a  Kullback–Leibler divergence loss (denoted as $\mathcal{L}_{KL}$) to them by regularizing the model to fairly capture the interest distributions of both heavy and cold users, which is formulated as follows:
\begin{equation}
    \mathcal{L}_{KL}=\sum_i D_{KL}(y_i,\hat{y}_i),
\end{equation}
where $y_i$ and $\hat{y}_i$ are the two kinds of click scores for the $i$-th training sample.
We denote the click prediction loss for recommendation model training as $\mathcal{L}_{Rec}$ (only applied to the $y$ predicted from original user data).
The overall training loss $\mathcal{L}$ is the summation of the two losses, i.e., $\mathcal{L}=\mathcal{L}_{Rec}+\mathcal{L}_{KL}$.

\section{Experiments on BigFair}

\begin{figure}[!t]
  \centering 
  \subfigure[Cold user.]{
      \includegraphics[width=0.7\linewidth]{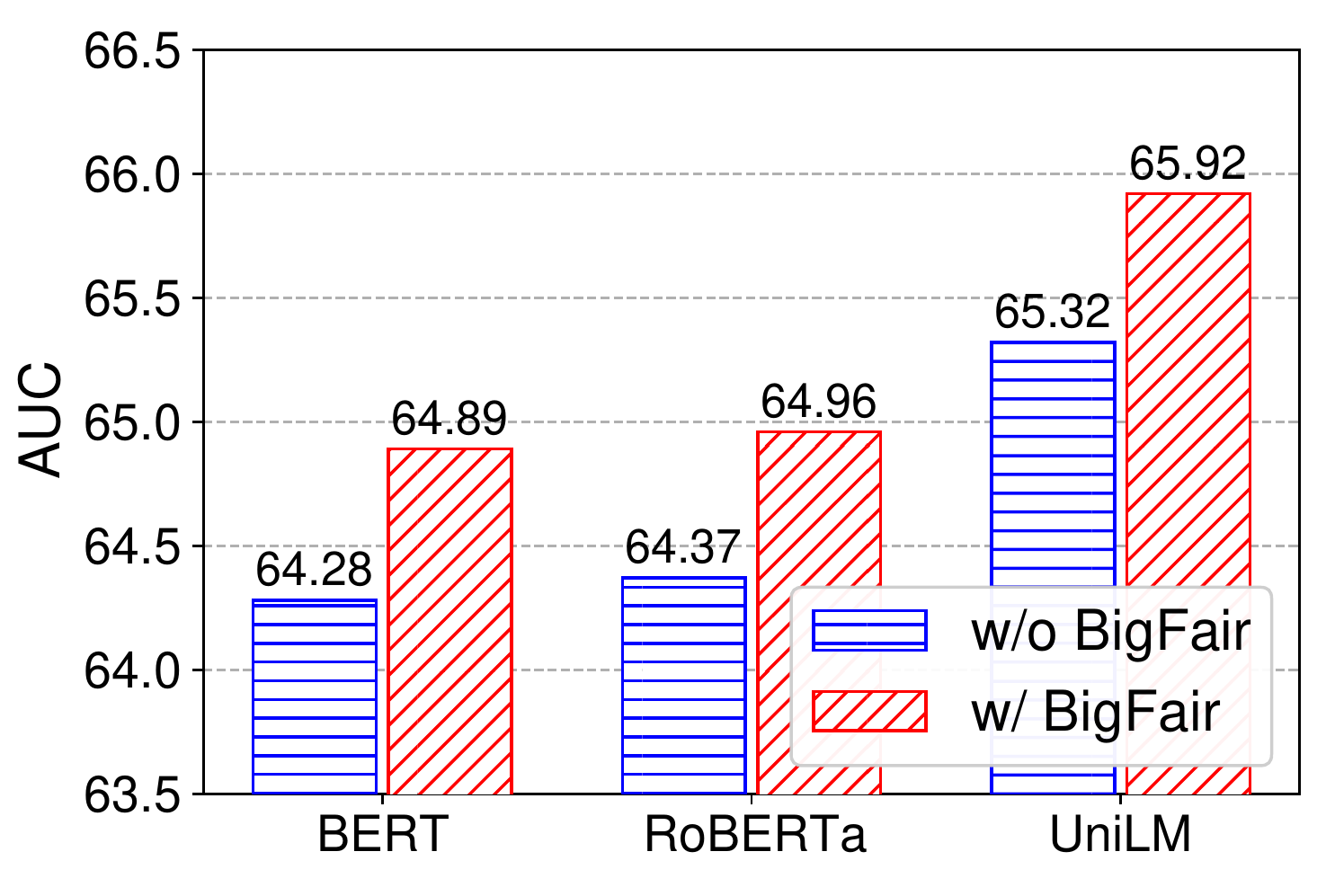} 
      }
        \subfigure[Heavy user.]{
      \includegraphics[width=0.7\linewidth]{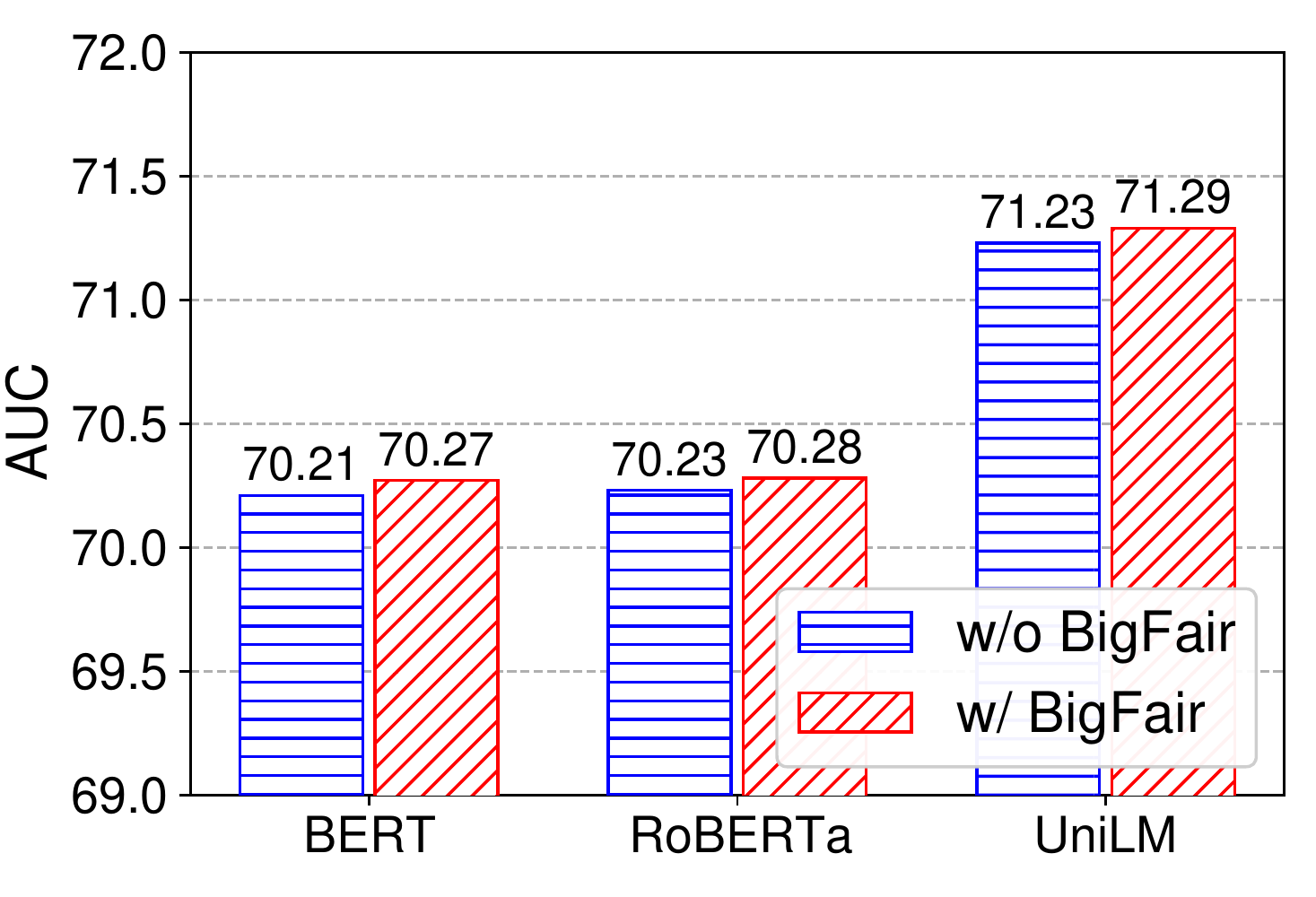} 
      }
  \caption{Effectiveness of \textit{BigFair} on \textit{MIND}.}\label{big}  
\end{figure}

\begin{figure}[!t]
  \centering 
      \includegraphics[width=0.7\linewidth]{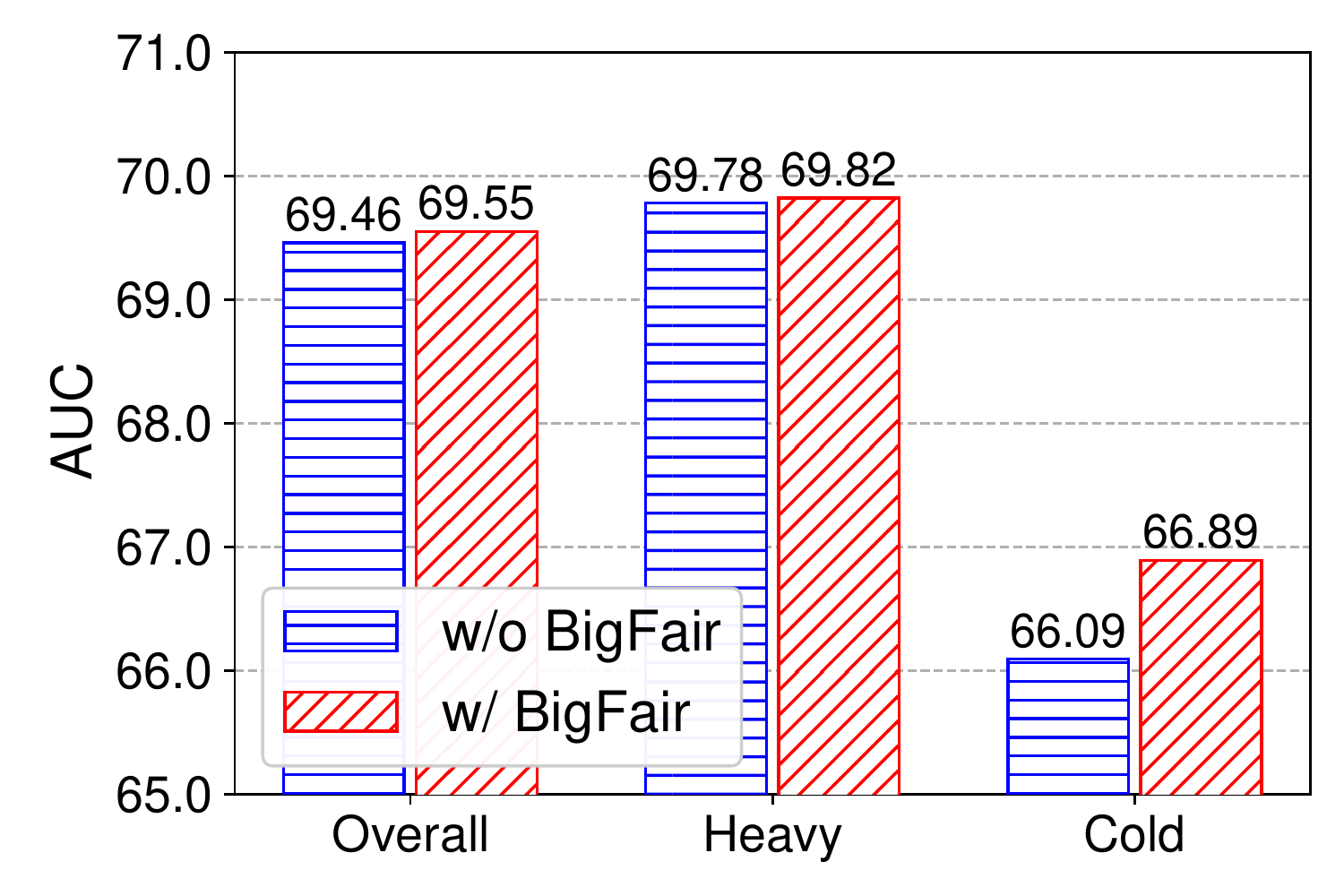} 
  \caption{Effectiveness of  \textit{BigFair} on \textit{Industry}.}\label{big2}  
\end{figure}

\begin{figure}[!t]
  \centering 
      \includegraphics[width=0.7\linewidth]{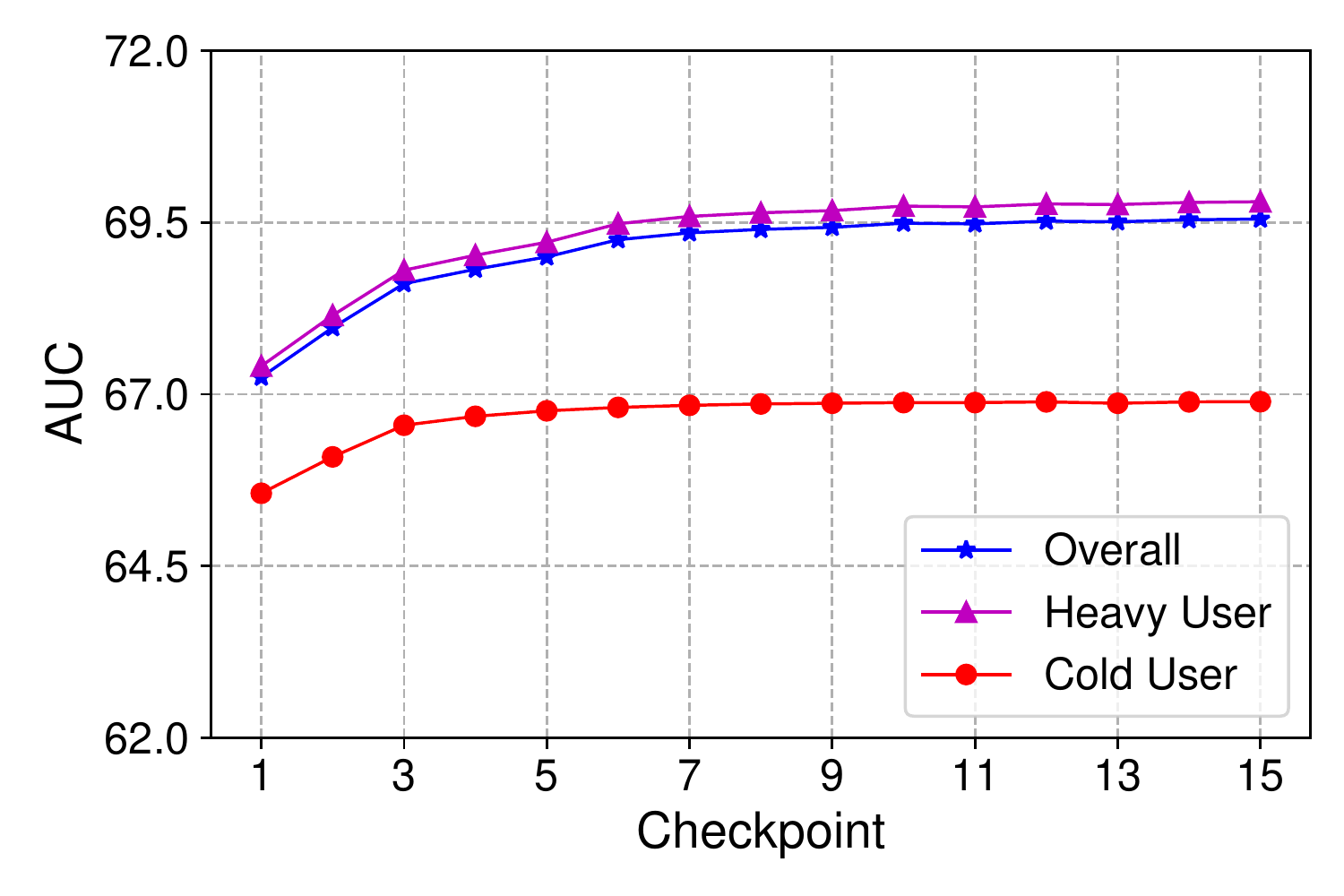} 
  \caption{Different checkpoint performance on \textit{Industry} of the model with \textit{BigFair}.}\label{big3}  
\end{figure}

\begin{figure}[!t]
  \centering 
      \includegraphics[width=0.7\linewidth]{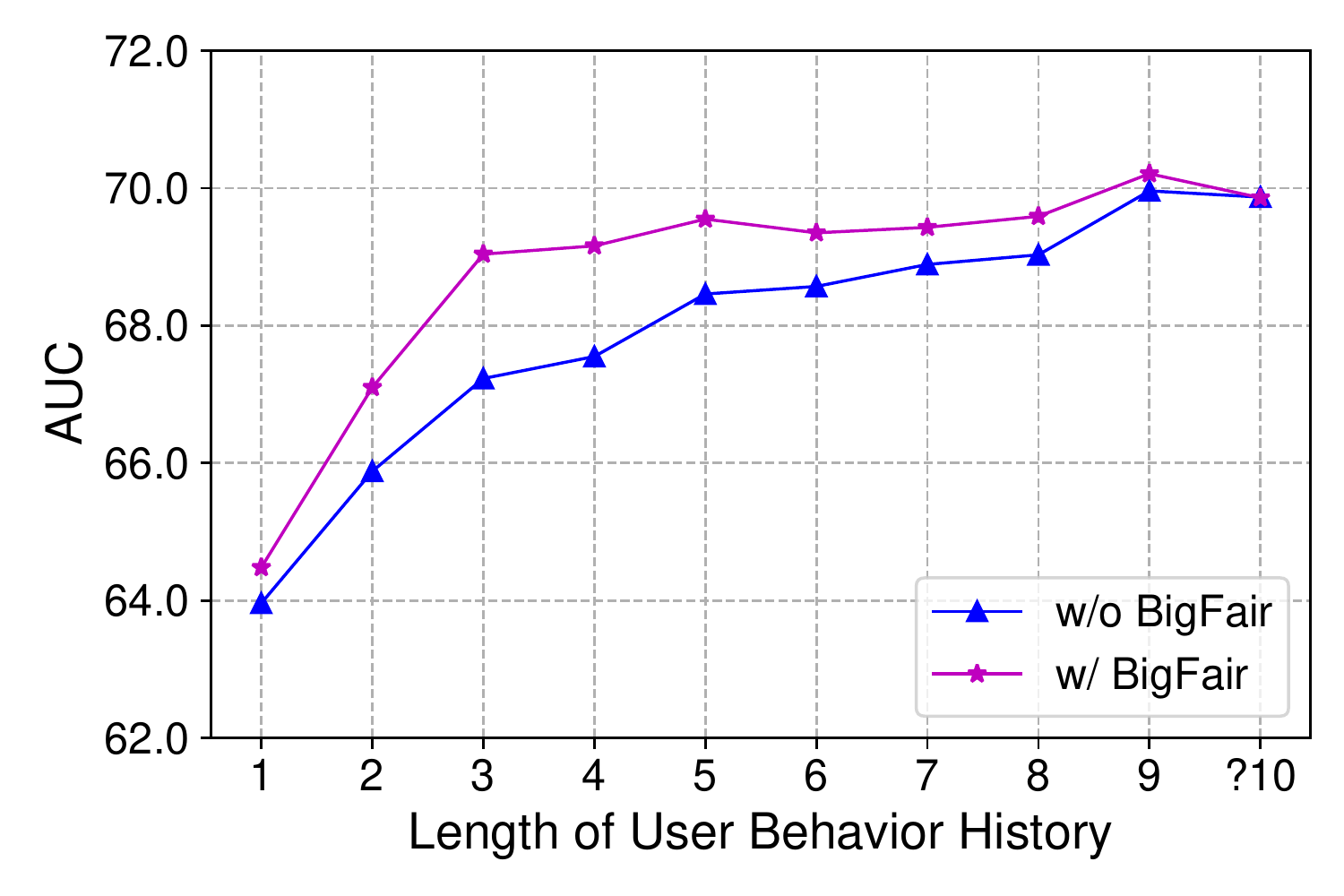} 
  \caption{Performance on users with different behavior history lengths on \textit{Industry}.}\label{len}  
\end{figure}

We further conduct experiments to validate the effectiveness of our proposed \textit{BigFair} method.
In our experiments, we set the behavior dropping ratio $P$ to 0.5.
We compare the performance of different big recommendation models (i.e., PLM-NR with different PLMs) and their variants combined with our \textit{BigFair} method.
The results on \textit{MIND} are shown in Fig.~\ref{big}.
We find that \textit{BigFair} can consistently improve the performance of different models on cold users (even better than the optimal models on cold users in Table~\ref{result1}), and meanwhile do not harm the results on heavy users.
We also compare the performance of PLM-NR (UniLM) with or without \textit{BigFair} on the \textit{Industry} dataset, as shown in Fig.~\ref{big2}.
The results  show that \textit{BigFair} can effectively improve the performance on cold users and can even bring slight improvements on heavy users.
We also illustrate the performance of different intermediate checkpoints of the model with  \textit{BigFair} in  Fig.~\ref{big3}.
It  shows that optimizing the overall performance of big recommendation models trained by our \textit{BigFair} method usually can benefit cold users.
Fig.~\ref{len} shows the performance of models with or without \textit{BigFair} on users with different activeness indicated by the lengths of behavior history.
We can see that \textit{BigFair} can effectively improve the performance on users with different activeness, especially on cold users with few behaviors.
These results indicate that our \textit{BigFair} method can effectively mitigate the unfairness of big recommendation models to cold users.

\section{Conclusion}\label{sec:Conclusion}

In this paper, we investigate the problem that whether big recommendation models are fair to cold users.
Through extensive experiments on two real-world datasets, we empirically demonstrate that optimizing the overall performance of big recommendation models will lead to notable unfairness to cold users.
To solve this issue, we propose a \textit{BigFair} approach based on self-distillation, which uses the model prediction on original user data as a virtual teacher to regularize  predictions on augmented data with randomly dropped behaviors.
In this way, the model is encouraged to fairly capture the interest distributions of both cold and heavy users.
Extensive experiments on two datasets show that \textit{BigFair} can effectively mitigate the unfairness of big models to cold users and meanwhile do not bring performance sacrifice to heavy users.

\bibliography{main}
\bibliographystyle{acl_natbib}
\appendix
\clearpage
\section{Appendix}

\subsection{Experimental Environment}

Our experimental environment is built on a cloud GPU cluster.
The version of Python is 3.6.9.
For experiments on \textit{MIND}, we use 4 Tesla V100 GPUs with 32GB memory.
For experiments on \textit{Industry}, we use 8 Tesla V100 GPUs.
We use Pytorch 1.9 to implement the models and experiments.

\subsection{Hyperparameter Settings}

The complete hyperparameter settings of our approach are listed in Table~\ref{hyper}.

\begin{table}[h]
\centering
\resizebox{1.0\linewidth}{!}{
\begin{tabular}{l|cc}
\Xhline{1pt}
\multicolumn{1}{c|}{\textbf{Hyperparameters}}& \textbf{Small Model} & \textbf{Big Model} \\ \hline
 token embedding dimension                     & 300  & 768             \\  
hidden dimension                    & 400 & 768   \\ 
attention head                      & 20  & 16   \\ 

Transformer layer                     & 1 & 12   \\  
dropout                                      & 0.2 & 0.2       \\
max title length                                    & 30& 30  \\ 
max history length                                    & 50   & 50  \\
$P$                                 & - & 0.5       \\
optimizer                                    & Adam& Adam       \\
learning rate                                 & 1e-4& 3e-6       \\
batch size                                   & 32  & 32/128      \\  
max epoch                                   & 3  & 2\\ 
\Xhline{1pt}
\end{tabular}
}
\caption{Hyperparameter settings.}\label{hyper}
\end{table}

\subsection{Training Details}

For small model training on \textit{MIND}, the time costs of NAML, LSTUR and NRMS are 2.8h, 3.6h, 3.3h, respectively.
For big model training on \textit{MIND}, the time cost is around 30h.
For big model training on \textit{Industry}, the maximum time cost is around 8 days.
We use 16-bit float accuracy to accelerate model training.
We employ early stopping mechanism with a interval of 30,000 iterations.

\subsection{Hyperparameter Analysis}

\begin{figure}[!t]
  \centering 
  \subfigure[\textit{MIND}.]{
      \includegraphics[width=0.8\linewidth]{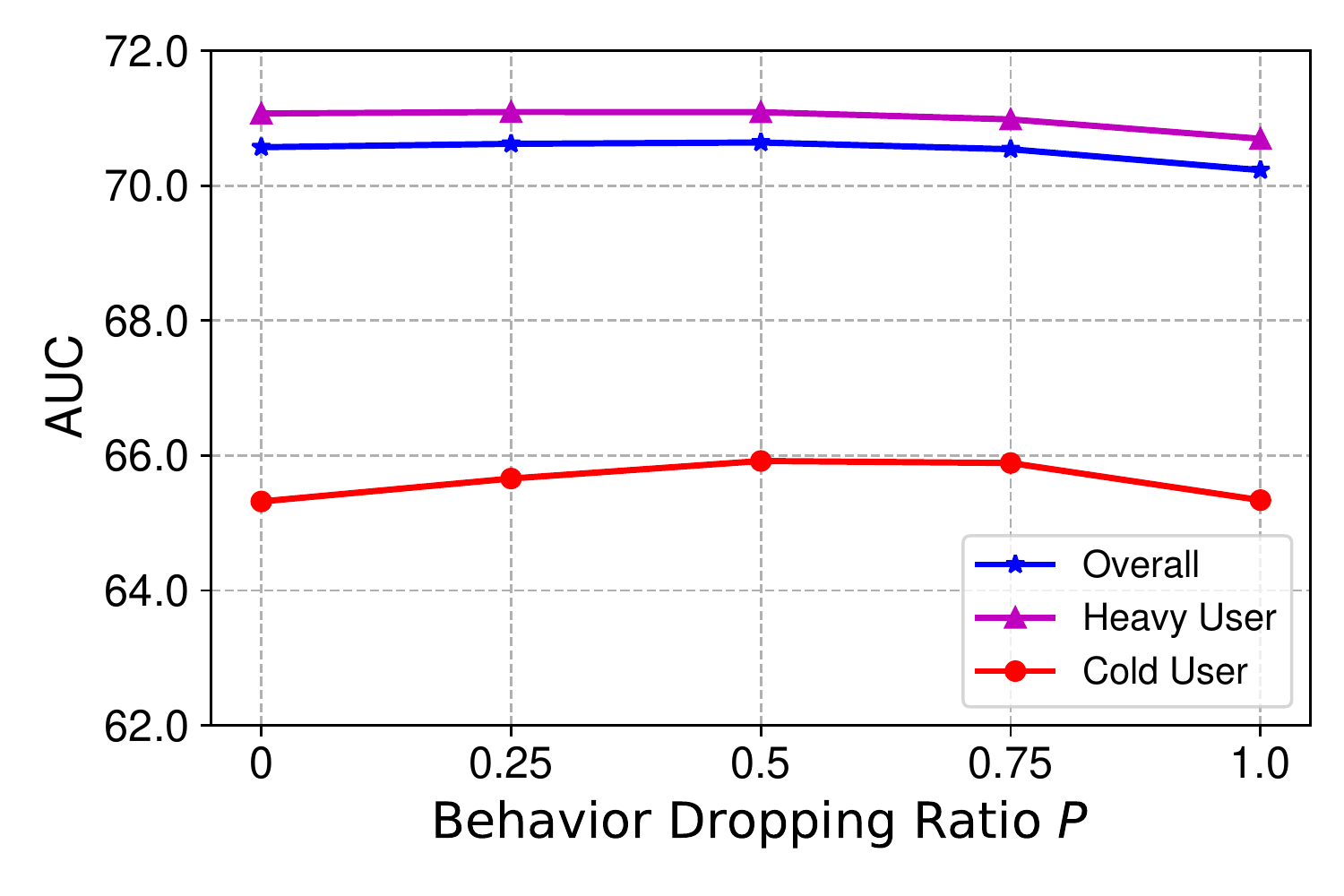} 
      }
        \subfigure[\textit{Industry}.]{
      \includegraphics[width=0.8\linewidth]{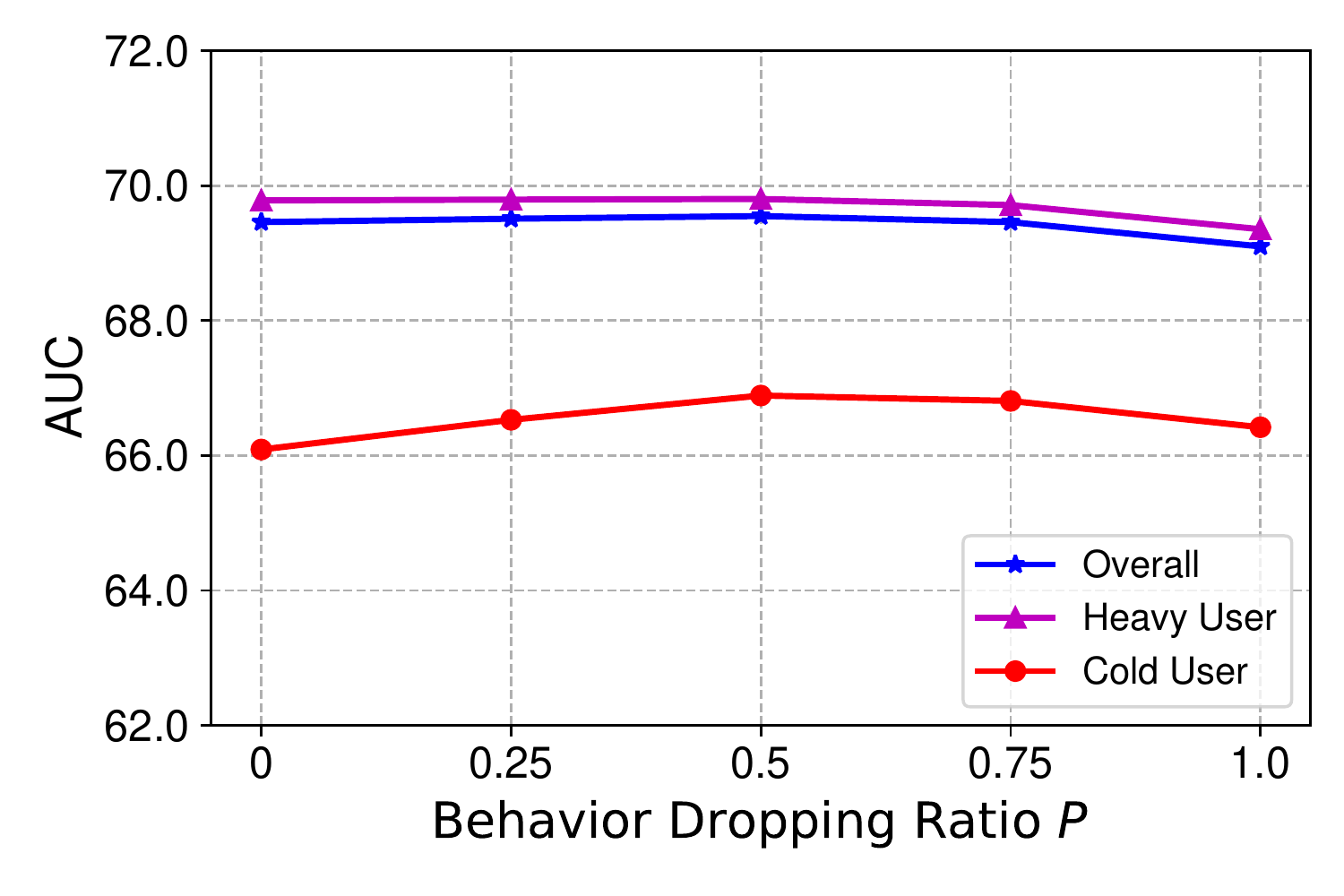} 
      }
  \caption{Influence of the behavior dropping ratio $P$.}\label{ratio}  
\end{figure}

We study the influence of a key hyperparameter in \textit{BigFair}, i.e., the behavior dropping ratio $P$ for generating augmented behavior sequences.
We vary the value of $P$ and compare the results of PLM-NR (UniLM) on both datasets, as shown in Fig.~\ref{ratio}.
We find the unfairness to cold users cannot be effectively addressed if $P$ is too small.
This is because the augmented user behavior sequence can also be long if $P$ is small, which is not suitable for simulating cold users.
However, when $P$ is too large, the performance also declines.
This is because when too many behaviors are dropped,  useful interest information is lost and the supervision signals can be noisy, which is also harmful for robust model training.
Thus, we choose $P=0.5$ that yields the best performance by dropping half of behaviors for augmenting behavior sequences.

\end{document}